\documentclass[twocolumn,american,english,aps]{revtex4}
\usepackage[T1]{fontenc}
\usepackage{amsmath}
\usepackage{amssymb}
\usepackage{graphicx}
\usepackage{esint}
\usepackage{bbm}
\usepackage{bm}

\makeatletter
\@ifundefined{textcolor}{}
{%
 \definecolor{BLACK}{gray}{0}
 \definecolor{WHITE}{gray}{1}
 \definecolor{RED}{rgb}{1,0,0}
 \definecolor{GREEN}{rgb}{0,1,0}
 \definecolor{BLUE}{rgb}{0,0,1}
 \definecolor{CYAN}{cmyk}{1,0,0,0}
 \definecolor{MAGENTA}{cmyk}{0,1,0,0}
 \definecolor{YELLOW}{cmyk}{0,0,1,0}
}

\makeatother

\usepackage{babel}
\begin{document}

\title{Floquet engineering of long-range p-wave superconductivity}

\author{M. Benito$^{1}$}
\email{m.benito@csic.es}
\author{A. G\'omez-Le\'on$^{1,2,3}$, V. M. Bastidas$^{4}$, T.
Brandes$^{4}$, and G. Platero$^{1}$}
\affiliation{$^{1}$Instituto de Ciencia de Materiales, CSIC, Cantoblanco, Madrid
E-28049, Spain\\
$^{2}$Department of Physics \& Astronomy, University of British Columbia,
Vancouver, BC, Canada V6T 1Z1\\
$^{3}$Pacific Institute of Theoretical Physics, Vancouver, BC V6T
1Z1, Canada\\
$^{4}$Institut f\"{u}r Theoretische Physik, Technische Universit\"{a}t Berlin,
Hardenbergstr. 36, 10623 Berlin, Germany}
\begin{abstract}
Floquet Majorana Fermions appear as steady states at the boundary of time-periodic
topological phases of matter. In this work, we theoretically study the main features of these exotic topological phases
in the periodically driven one-dimensional  Kitaev model. By controlling the ac
fields, we can predict new topological phase transitions that
should give rise to signatures of Majorana states in experiments.
Moreover, the knowledge of the time-dependence of these Majorana states
allows one to manipulate them. Our work contains a complete analysis of
the monochromatic driving in different frequency regimes. 
\end{abstract}

\maketitle
\section{Introduction\label{SectionI}}
In the last years, phases of matter with topological origins have received intense attention in solid state physics~\cite{Kane2005,Bernevig2006,Fu2007,Fu2007a,Hsieh2008,Konig2007,Thouless1982}. In topological superconductors these phases result in elementary excitations
with non-abelian statistics~\cite{Kitaev2001,Alicea2012}. The bulk-boundary
correspondence shows how differences between bulk topologies give
rise to edge states localized at the boundaries~\cite{Hasan2010}. 

Driven systems constitute a fruitful arena to study topological states of matter  because they
exhibit topological features that are richer than those of time-independent
systems. Of particular interest are periodically driven systems, which
have many close analogies with static systems~\cite{Inoue2010,Kitagawa2010,Kitagawa2010a,Lindner2011,Rudner2013,Gomez-Leon2013}.
Floquet eigenstates and quasienergies are similar to eigenstates and
energies of static systems, but the periodicity of the quasienergy
spectrum introduces new features unique to periodically driven systems~\cite{Jiang2011,Kitagawa2010,Rudner2013,Kitagawa2012,Kundu2013,Tong2013}.
In fact, the complete characterization of the topology of a time-periodic Hamiltonian
requires the search for new topological invariants~\cite{Asboth2012,Asboth2013,Tarasinski2014,Kitagawa2010}.

In this work, we propose an analytic approach to describe the topological phases
of a driven chain of spinless fermions with $p$-wave superconductivity, i.e., the one-dimensional
Kitaev model~\cite{Kitaev2001}. The end states emerging in driven
topological superconductors are called Floquet Majorana Fermions, because they are the analogous counterpart for Majorana Fermions in static systems ~\cite{Mourik2012,Rokhinson2012,Das2012,Kitaev2001,Deng2012,Oreg2010,Lutchyn2010,PhysRevB.86.140503}.
The characterization of these excitations allows one to design protocols
for their manipulation, which is potentially relevant for braiding
operations, which are essential for fault-tolerant quantum computation~\cite{Liu2013}. Furthermore, periodic driving opens a new avenue to detect
these elusive particles~\cite{Kundu2013,Wang2014}. 

In the last years, some works have addressed the effect of ac driving fields
in topological superconductors. Most of them are restricted to the
high-frequency regime. Those addressing lower frequencies are mainly focused
on numerical treatments~\cite{Wu2013,Liu2013,Jiang2011,Reynoso2013}. A more complete
analysis was done in the cases of periodically kicked systems~\cite{Thakurathi2013} and step-like periodic pulses~\cite{Tong2013}, including
the definition of new topological invariants, while the harmonic driving
is treated just numerically~\cite{Thakurathi2013}.  

In contrast to previous works, in this paper we develop a complete analysis of monochromatic driving for different
frequency regimes, which is more feasible in experiments. Our analytical treatment allows us to characterize different
topological phases by means of effective Hamiltonians in rotated reference frames.
Furthermore, we address different ways to drive the Kitaev chain giving rise to a variety of topological quantum phase transitions (TQPTs), which could be experimentally tested.

The correspondence between the Kitaev chain and the one-dimensional transverse Ising model~\cite{Niu2012} is used mainly as a mathematical tool to simplify the analysis
and to set a different framework to probe the phase transitions. In this sense, our results not only have relevance in the field of topological states of matter, but also provide
insight in the study of quantum magnetism under nonequilibrium situations~\cite{Dziarmaga2005,ADas2010,Bastidas2012,Santoro2012}. 

The simple characterization of the driven system for arbitrary frequencies
by means of rotations of frame is the most important result of our work.
This allows us to obtain the wave function of the Majorana end states in
an easy way and to understand the role of the quasienergies in the TQPTs.
Moreover, we show that the driving protocols allows one to manipulate the 
effective interactions between different neighbors, generating effective models that are difficult to implement in 
time-independent systems. Apart from this, we establish a connection between the effective interactions generated in the Kitaev model under the effect of driving and the magnetic interactions in the Ising model.

In section~\ref{SectionII} we introduce the model and describe the main tools used in this paper.
In particular, the behavior of Floquet states and the Magnus Expansion
under a rotation of the reference frame is studied in detail. In section~\ref{SectionIII} we analyze the case
of an ac driven 
chemical potential. In this section
we present a thorough discussion of the methodology
used to determine the TQPTs, based on effective Hamiltonians in different frames. In sections~\ref{SectionIV} and~\ref{SectionV} we consider different 
driving protocols and discuss the emergence of exotic phases. In particular, we discuss the effective long-range interactions arising under the control of the tunneling amplitude.

\section{Model and Tools \label{SectionII}}
The system we analyze consists of a chain with $N\gg1$ sites.
Each site $j$ can be either empty or occupied by a spinless fermion
$f_{j}$. It consists in the driven version of the one-dimensional 
Kitaev model~\cite{Kitaev2001}
\begin{align}
\label{eq:kitaev} 
H(t)=&\frac{\mu(t)}{2}\sum_{j=1}^{N}\left(2f_{j}^{\dagger}f_{j}-1\right)-\frac{w(t)}{2}\sum_{j=1}^{N} \left(f_{j}^{\dagger}f_{j+1}+\text{h.c.}\right)
\nonumber\\ &
-\frac{\Delta(t)}{2}\sum_{j=1}^{N}\left(f_{j}^{\dagger}f^{\dagger}_{j+1}+\text{h.c.}\right) \ ,
\end{align}
where $\mu(t)$ is the chemical potential, $w(t)$ the tunneling and $\Delta(t)$ the BCS superconducting pairing between nearest neighbors 
in the presence of driving. In the case of periodic boundary conditions $f_{N+1}=f_{1}$, we can use a discrete Fourier transformation to obtain the bulk Hamiltonian
\begin{eqnarray}
      \label{eq:k}
     H_{k}(t)& = & 
     \begin{pmatrix}
     \mu(t)-w(t)\cos k & -i\Delta(t)\sin k\\
     i\Delta(t)\sin k & -\mu(t)+w(t)\cos k
     \end{pmatrix} \nonumber
     \\
     &=&[\mu(t)-w(t)\cos k]\sigma^{z}_k+\Delta(t)\sin k\ \sigma^{y}_k 
     \ ,
\end{eqnarray}
where $\sigma^{\lambda}_k$ for $\lambda\in\{x,y,z\}$ are the Pauli matrices in Nambu space. Correspondingly, we can write Eq.~\eqref{eq:kitaev} as $H(t)=\sum_{k>0}\Psi^{\dagger}_{k} H_{k}(t)\Psi_{k}$, where $\Psi^{\dagger}_{k}=(f_{k}^{\dagger}, f_{-k})$ and $f_{k}$ are fermionic operators in reciprocal space~\cite{Lieb1961}.

The undriven model, i.e., $\mu(t)=\mu_0$, $w(t)=w_0$ and $\Delta(t)=\Delta_0$, undergoes a TQPT~\cite{Kitaev2001}. Given that $\Delta_0>0$, the system exhibits a topologically nontrivial phase when $\mu_{0}<w_{0}$ and  a topologically trivial behavior as long as $\mu_{0}>w_{0}$. 
The Hamiltonian has particle-hole and time-reversal symmetry ~\cite{Niu2012,Tewari2012} and therefore the different topological phases can be classified by means of the value of a bulk $\mathbb{Z}$ topological invariant, which corresponds to the winding number 

\begin{equation}
      \label{eq:WindingNumber}
            W=\frac{1}{2\pi}\int_{-\pi}^{\pi}d \varphi_k
       \ ,
\end{equation}
where $\tan\varphi_k=\Delta_{0}\sin k\ (\mu_{0}-w_{0}\cos k)^{-1}$. The winding number
is $W=1$ in the nontrivial phase, and $W=0$ in the trivial one.
A chain in the nontrivial phase with open boundary conditions exhibits Majorana modes localized at the ends~\cite{Kitaev2001}.

The Hamiltonian of the driven $\text{XY}$ model in an external transverse field 
\begin{equation}
      \label{eq:ising-1}
            H(t)=-\frac{1}{2}\sum_{j=1}^{N}\left[\mu(t)\sigma_{j}^{x}-J_{z}(t)\sigma_{j}^{z}\sigma_{j+1}^{z}-J_y(t)\sigma_{j}^{y}\sigma_{j+1}^{y}\right]
      \ 
\end{equation}
can be exactly mapped onto the Kitaev Hamiltonian in Eq.~\eqref{eq:kitaev}
by means of a Jordan-Wigner transformation~\cite{Lieb1961}. 
The time-dependent anisotropies are related to the tunneling and the superconducting gap as: $J_z(t)=[w(t)+\Delta(t)]/2$ and $J_y(t)=[w(t)-\Delta(t)]/2$. 
In the time-independent case, there is a correspondence between the 
magnetic phases of the spin system and the topological phases of the 
fermion model, i.e., the paramagnetic phase is related to the trivial 
phase, and the ferromagnetic phase corresponds to the nontrivial phase~\cite{PhysRevB.88.165111}. 
Under monochromatic driving new magnetic phases arise, corresponding to new topological phases in the fermionic system.

\subsection{Floquet Theory under rotation of the reference frame}
We analyze the effect of a periodic time-dependence of
the Hamiltonian in Eq.~\eqref{eq:kitaev}. In this case $H(t+T)=H(t)$ (with $T=2\pi/\omega$
the period of the driving), therefore Floquet theory is applicable~\cite{Shirley1965,Sambe1973,Platero2004}.
By using the Floquet states $|\psi_{\nu}(t)\rangle=e^{-i\epsilon_{\nu} t}|\phi_{\nu}(t)\rangle$ the 
time-dependent Schr\"{o}dinger equation becomes an eigenvalue equation for the Floquet modes $|\phi_{\nu}(t+T)\rangle=|\phi_{\nu}(t)\rangle$, referred to as the Floquet equation
\begin{equation}
      \label{eq:FloquetEquation}
      \left[H(t)-i\partial_{t}\right]|\phi_{\nu}(t)\rangle=\epsilon_{\nu}|\phi_{\nu}(t)\rangle
      \ . 
\end{equation}
The operator $\mathcal{H}(t)=H(t)-i\partial_{t}$
is the Floquet Hamiltonian, its eigenvalues are the quasienergies $\epsilon_{\nu}$ and the eigenvectors $|\phi_{\nu}(t)\rangle$ are the Floquet modes. The index $\nu$ corresponds to the band index. 
The quasienergies $\epsilon_{\nu}$ are not uniquely defined~\cite{Shirley1965,Sambe1973,Platero2004}. Therefore,  we restrict them to the first Brillouin 
zone $-\omega/2\leq\epsilon_{\nu}\leq\omega/2$.

In order to simplify the resolution of the Floquet equation one should find an appropriate rotating frame in which a simple effective Hamiltonian can be defined, as we describe below.
First of all we will clarify the effect of a rotation of frame
in the Floquet formalism. In the rotating frame, given by the unitary
transformation ${\cal S}(t)$,
the Floquet equation becomes $\left[\tilde{H}(t)-i\partial_{t}\right]|\tilde{\phi}_{\nu}(t)\rangle=\epsilon_{\nu}|\tilde{\phi}_{\nu}(t)\rangle$,
where the Hamiltonian in the new frame is  
\begin{equation}
      \label{eq:DefHamRot}
            \tilde{H}(t)={\cal S}^{\dagger}(t)H(t){\cal S}(t)-i{\cal S}^{\dagger}(t)\dot{{\cal S}}(t)
      \ ,
\end{equation}
and $|\tilde{\phi}_{\nu}(t)\rangle={\cal S}^{\dagger}(t)|\phi_{\nu}(t)\rangle$.
As a consequence,  the quasienergies do not change,
but the periodicity of the Floquet modes does
\begin{equation}
\label{eq:floquetframe1}
\begin{split}
|\tilde{\phi}_{\nu}(t+T)\rangle & =  {\cal S}^{\dagger}(t+T){\cal S}(t)|\tilde{\phi}_{\nu}(t)\rangle\ .
\end{split}
\end{equation}
The one-period time-evolution operator makes a state evolve as $U(T,0)|\psi(0)\rangle=|\psi(T)\rangle$. Apart from Eq.~\eqref{eq:FloquetEquation}, the quasienergies can also be obtained from the eigenvalue equation $U(T,0)|\phi_{\nu}(0)\rangle=e^{-i\epsilon_{\nu} T}|\phi_{\nu}(0)\rangle$.
If we are working in a rotated frame 
we can use Eq.~\eqref{eq:floquetframe1} to obtain the eigenvalue equation for the one-period evolution operator
\begin{equation}
\tilde{U}(T,0)|\tilde{\phi}_{\nu}(0)\rangle=e^{-i\epsilon_{\nu} T}{\cal S}^{\dagger}(T){\cal S}(0)|\tilde{\phi}_{\nu}(0)\rangle\label{eq:Floquetframe2}
  \ .
\end{equation}
As a consequence of this, the relation between quasienergies $\epsilon_{\nu}$ and eigenvalues
of $\tilde{U}(T,0)$ depends on the matrix ${\cal S}$. A common method to study time-dependent systems is to calculate the eigenvalues of the one-period evolution operator~\cite{Shirley1965,Sambe1973,Platero2004}. Eq.~\eqref{eq:Floquetframe2}  shows the relationship between these eigenvalues in an arbitrary reference frame (given by ${\cal S}$) and the quasienergies.

\subsection{Magnus expansion for effective Hamiltonians}
Once we are in the most suitable reference frame, we define a time-independent
effective Hamiltonian $\tilde{H}_{\text{eff}}$ such that 
\begin{equation}
\tilde{U}(T,0)=\exp\left( -i\tilde{H}^{\text{eff}}T\right) \ .\label{eq:time-ordered}
\end{equation}
with $\tilde{H}(t)$ periodic in time, the Fourier decomposition
$\tilde{H}(t)=\sum_{p}e^{ip\omega t}\tilde{H}_{p}$ allows us to write
a power series expansion in $\frac{1}{\omega}$ for the time-independent Hamiltonian, referred to as the Magnus expansion~\cite{Blanes2009,Lopez2013}
\begin{equation}
\label{eq:magnus-1}
\begin{split}
  \tilde{H}^{\text{eff}} =& \tilde{H}_{0}+\frac{1}{\omega}\left[\tilde{H}_{0},\tilde{H}_{1}\right]-\frac{1}{\omega}\left[\tilde{H}_{0},\tilde{H}_{-1}\right]\\
&-\frac{1}{\omega}\left[\tilde{H}_{-1},\tilde{H}_{1}\right]+\ldots
\end{split}
\end{equation}
with an infinite number of terms.
This series is only useful if it converges with a finite number of
terms. The convergence is given by the condition $\int_{0}^{T}\left\Vert \tilde{H}(t)\right\Vert dt<\pi$, 
where $\left\Vert \tilde{H}(t)\right\Vert $ is the Euclidean norm
of the Hamiltonian as it is discussed in Refs.~\cite{Blanes2009,Lopez2013}. At this point, the choice of the frame of reference is relevant,
since the Hamiltonian in different frames $\tilde{H}(t)$ has different
convergence conditions.

The previous explanations are valid for general time-periodic Hamiltonians. In particular, the expansion Eq.~\eqref{eq:magnus-1} can be used for a Bogoliubov-de-Gennes Hamiltonian in Nambu space. In this case,  the transformation into the rotating frame factorizes as $\mathcal{S}=\bigotimes_{k>0}\mathcal{S}_{k}$. For a given $k$ we can use Eq.~\eqref{eq:DefHamRot} to obtain the transformation of Hamiltonian Eq.~\eqref{eq:k} into the rotating frame  $\tilde{H}_{k}(t)=\mathcal{S}^{\dagger}_{k}H_{k}(t)\mathcal{S}_{k}-i\mathcal{S}^{\dagger}_{k}\dot{{\cal S}_k}(t)$. 
Therefore, the convergence condition reads
\begin{equation}
      \label{eq:convNambu}
            \int_{0}^{T}\left\Vert \tilde{H}_{k}(t)\right\Vert dt<\pi
      \ .
\end{equation}
In Nambu space, Eq.~\eqref{eq:Floquetframe2} implies the eigenvalue problem for the Floquet modes
\begin{equation}
\tilde{U}_{k}(T,0)|\tilde{\phi}_{\nu,k}(0)\rangle=e^{-i\epsilon_{\nu,k} T}{\cal S}_{k}^{\dagger}(T){\cal S}_{k}(0)|\tilde{\phi}_{\nu,k}(0)\rangle\label{eq:FloquetframeNambu}
  \ ,
\end{equation}
where $\epsilon_{\nu,k}$ is the $\nu$-th band quasienergy dispersion, $|\tilde{\phi}_{\nu}(0)\rangle=\bigotimes_{k>0}|\tilde{\phi}_{\nu,k}(0)\rangle$ and $\tilde{U}(T,0)=\bigotimes_{k>0}\tilde{U}_{k}(T,0)$.

\section{Driving the Chemical Potential\label{SectionIII}}
We focus here on the study of the topological properties of the Kitaev model
with time-dependent chemical potential. Therefore, we assume $w(t)=w_0$, $\Delta(t)=\Delta_0$ and $\mu(t)=\mu_{0}+\frac{\mu_{1}}{2}\cos\omega t$,
where $\mu_{0}$ is a constant term and $\frac{\mu_{1}}{2}$ is the amplitude
of the driving. 

As the Kitaev model is described by a Bogoliubov-de-Gennes Hamiltonian the solutions give the spectrum of excitations. Due to the particle-hole symmetry these excitations come in pairs such that creating an excitation with energy $E$ is equivalent to annihilating the excitation with energy $-E$. It means $\gamma_{E}^{\dagger}|GS\rangle=\gamma_{-E}|GS\rangle$, where $|GS\rangle$ is the vaccum of excitations fulfilling $\gamma_{E}|GS\rangle=0\ \forall \  E>0$.
Given this simmetry, a zero-energy excitation will fulfill the Majorana
condition $\gamma_{0}=\gamma_{0}^{\dagger}$. In the driven
case, excitations come also in pairs $\gamma_{\epsilon}(t)=\gamma_{-\epsilon}^{\dagger}(t)$, where $\epsilon$ is the quasienergy. 
Due to the periodicity of the quasienergies,
not only $\epsilon=0$, but also $\epsilon=\pm\frac{\omega}{2}$ excitations
fulfill the Majorana condition $\gamma_{0,\frac{\omega}{2}}=\gamma_{0,\frac{\omega}{2}}^{\dagger}$ ~\cite{Jiang2011}. Therefore
 a complete phase diagram has to take into
account possible closings of the quasienergy spectrum at quasienergies $\epsilon=\pm\frac{\text{\ensuremath{\omega}}}{2}$,
as well as $\epsilon=0$. Furthermore, both quasienergy gaps can support
Majorana end states 
and the topological phase is characterized by two 
$\mathbb Z$  topological invariants
($\mathbb Z \times \mathbb Z $). 
\begin{figure}
\includegraphics[width=0.7\columnwidth]{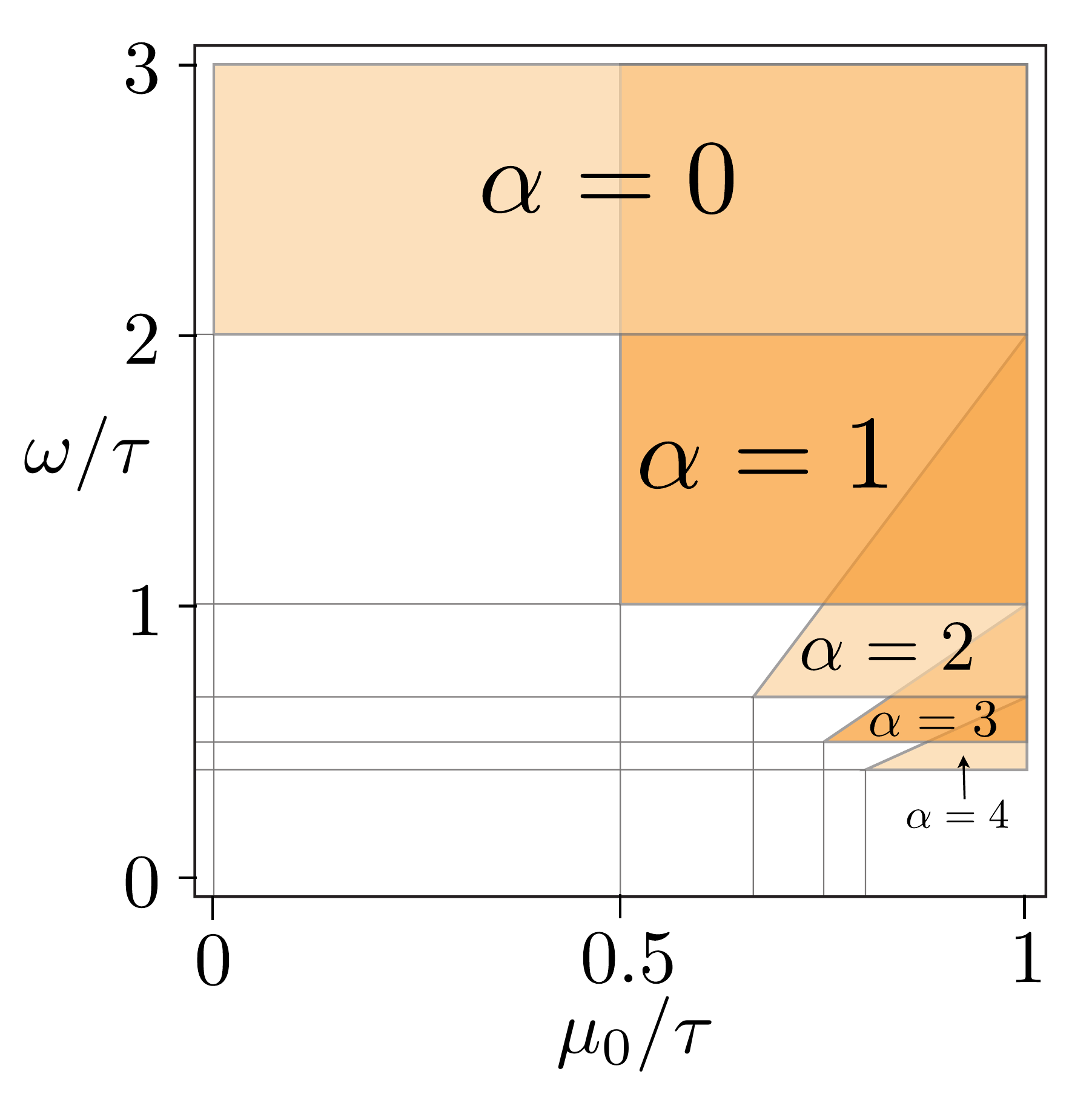}
\caption{Regions of convergence of the Magnus expansion of $\tilde{H}_{k}^{\alpha}(t)$ 
for $\alpha\in\{0,1,2,3,4\}$ as a function of $\mu_{0}$ and $\omega$. In order to obtain a convergent Magnus expansion, the Hamiltonian $\tilde{H}_{k}^{\alpha}(t)$ has to fulfill the convergence condition Eq.~\eqref{eq:convNambu} for all the values of $k$. 
At high frequency $\omega>2\tau$,
the series of $\tilde{H}_{k}^{0}(t)$ converges. For $\omega<2\tau$ the
successive $\tilde{H}_{k}^{\alpha}(t)$ have convergent series for some
values of $\mu_{0}$. We consider a fixed bandwidth $\tau=\mu_{0}+w_{0}$.\label{fig:convergence1}}
\end{figure}
In the following, we present a method based on reference frame transformations
to find the topological phase diagram of the Hamiltonian Eq.~\eqref{eq:kitaev}. Moreover,
our method provides the wave function of the Majorana
excitations.
\subsection{Reference frame choice}
Let us consider first the convergence of the Magnus expansion in the laboratory frame $\tilde{H}(t)=H(t)$. We calculate the first harmonics of Hamiltonian Eq.~\eqref{eq:k}:
\begin{equation}
      \label{eq:HarLab}
      \begin{split}
            H_{k,0}&=\frac{1}{T}\int_{0}^{T}dt'H_{k}(t')
             \\
            &=(\mu_{0}-w_{0}\cos k)\sigma^{z}_k+\Delta_{0}\sin k\ \sigma^{y}_k 
            \ , \\
            H_{k,\pm1}&=\frac{1}{T}\int_{0}^{T}dt'H_{k}(t')e^{\mp i\omega t}=\frac{\mu_{1}}{4}\sigma^{z}_k 
            \ .
             \end{split}
            \end{equation}
Regardless of the value of the quasimomentum $k$, the first term of the Magnus expansion $H_{k,0}$ is already a good approximation if the frequency is much larger than the bandwidth
$\tau=\mu_{0}+w_{0}$ and the driving amplitude $\mu_{1}$.
We will show below that by means of a rotation of frame the convergence regions can be increased. 

We work with a whole family of rotating frames given by the transformations
$S_{k,\alpha}^{\dagger}(t)=e^{i\theta_{\alpha}(t)\sigma^{z}_k}$ with $\theta_{\alpha}(t)=\frac{\alpha\omega}{2}t+\frac{\mu_{1}}{2\omega}\sin\omega t$ for 
$\alpha\in\{0,\pm1,\pm2...\}$. For $\alpha=0$ we obtain the transformation into the interaction picture. The Hamiltonians in the rotating frame are
\begin{align}
      \label{eq:Hm}
            \tilde{H}^{\alpha}_{k}(t)&=\left(\mu_{0}-\frac{\alpha\omega}{2}-w_{0}\cos k\right)\sigma^{z}_{k} \\
            & -i\Delta_{0}\sin k\  e^{2i\theta_{\alpha}(t)}\sigma^{+}_{k}
            +i\Delta_{0}\sin k\  e^{-2i\theta_{\alpha}(t)}\sigma^{-}_{k} 
            \nonumber \ .
\end{align}
Given that ${\cal S}_{k,\alpha}^{\dagger}(T){\cal S}_{k,\alpha}(0)=e^{i\alpha\pi\sigma^{z}_k}$, Eq.~\eqref{eq:FloquetframeNambu} becomes 
\begin{equation}
      \label{eq:EVEqNambu}
      \tilde{U}_{k}^{\alpha}(T,0)|\tilde{\phi}_{\nu,k}(0)\rangle=e^{-i\epsilon_{\nu,k} T}e^{i\alpha\pi\sigma^{z}_k}|\tilde{\phi}_{\nu,k}(0)\rangle
      \ .
\end{equation}
This leads to the eigenvalue equation for the effective bulk Hamiltonian in Eq.~\eqref{eq:time-ordered}
\begin{equation}
\tilde{H}_{k}^{\text{eff},\alpha}|\tilde{\phi}_{\nu,k}(0)\rangle=\left(\epsilon_{\nu,k}-\frac{\alpha\omega}{2}\sigma^{z}_{k}\right)|\tilde{\phi}_{\nu,k}(0)\rangle\ ,\label{eq:shift}
\end{equation}
which implies that the quasienergies and the eigenvalues of the effective
Hamiltonian are related by a $\frac{\alpha\omega}{2}$ shift. Due to the periodicity of the quasienergies, this shift is relevant just in the case of odd values of $\alpha$.

For a given $\alpha$ the transformed Hamiltonian of Eq.~\eqref{eq:Hm} has different regions
of convergence, determined  by the condition in Eq.~\eqref{eq:convNambu}. The Fourier components $\tilde{H}_{k,p}^{\alpha}\equiv\frac{1}{T}\int_{0}^{T}dt'\tilde{H}_{k}^{\alpha}(t')e^{- ip \omega t}$ of $\tilde{H}_{k}^{\alpha}(t)$ are given by
\begin{align}
      \label{eq:Hp}
           & \tilde{H}_{k,p}^{\alpha}=\left(\mu_{0}-\frac{\alpha\omega}{2}-w_{0}\cos k\right)\delta_{p,0}\sigma^{z}_{k} \\
            & -i\Delta_{0}\sin k\ {\cal J}_{p-\alpha}\left(\frac{\mu_{1}}{\omega}\right)\sigma^{+}_{k}
            +i\Delta_{0}\sin k\ {\cal J}_{-p-\alpha}\left(\frac{\mu_{1}}{\omega}\right)\sigma^{-}_{k}
            \nonumber \ ,
\end{align}
where ${\cal J}_{n}(x)$ is the $n$th-order Bessel function. We remark
that the eigenvalues of the Hamiltonian in Eq.~\eqref{eq:Hm} do not depend
on the amplitude $\mu_{1}$. Then, the convergence condition Eq.~\eqref{eq:convNambu}
is $\mu_{1}$-independent.  Fig.~\ref{fig:convergence1} depicts the
regions of convergence of the Magnus expansion for successive $\tilde{H}_{k}^{\alpha}(t)$.
The series of $\tilde{H}_{k}^{0}(t)$ converges in the high frequency regime $\omega>2\tau$. For $\omega<2\tau$ the
successive Hamiltonians $\tilde{H}_{k}^{\alpha}(t)$ have convergent series
for different values of $\mu_{0}$.

By obtaining a frame where the Hamiltonian has a convergent Magnus series,
we are able to get an approximation of the full quasienergy spectrum from the eigenvalues of the time-independent Hamiltonian of  Eq.~\eqref{eq:magnus-1}. Moreover, the effective Hamiltonian
allows a simple description of the TQPTs. Within the regions of convergence,
the effective Hamiltonian can be approximated by the zeroth-order term of Eq.~\eqref{eq:magnus-1}. 
Thereby, we can use Eq.~\eqref{eq:Hp} with $p=0$ as zeroth-order effective Hamiltonian
\begin{equation}
      \label{eq:Heffalpha}
      \tilde{H}_{k}^{\text{eff},\alpha}=\left(\mu_{\text{eff}}-w_0\cos k \right)\sigma^{z}_{k}+
      \Delta_{\text{eff}}\sin k\    \sigma^{y}_{k}
      \ ,
\end{equation}
which is an effective Kitaev model with $\mu_{\text{eff}}=\mu_{0}-\frac{\alpha\omega}{2}$ and $\Delta_{\text{eff}}=\Delta_{0}{\cal J}_{-\alpha}\left(\frac{\mu_{1}}{\omega}\right)$.
\begin{figure}
\includegraphics[width=1.\columnwidth]{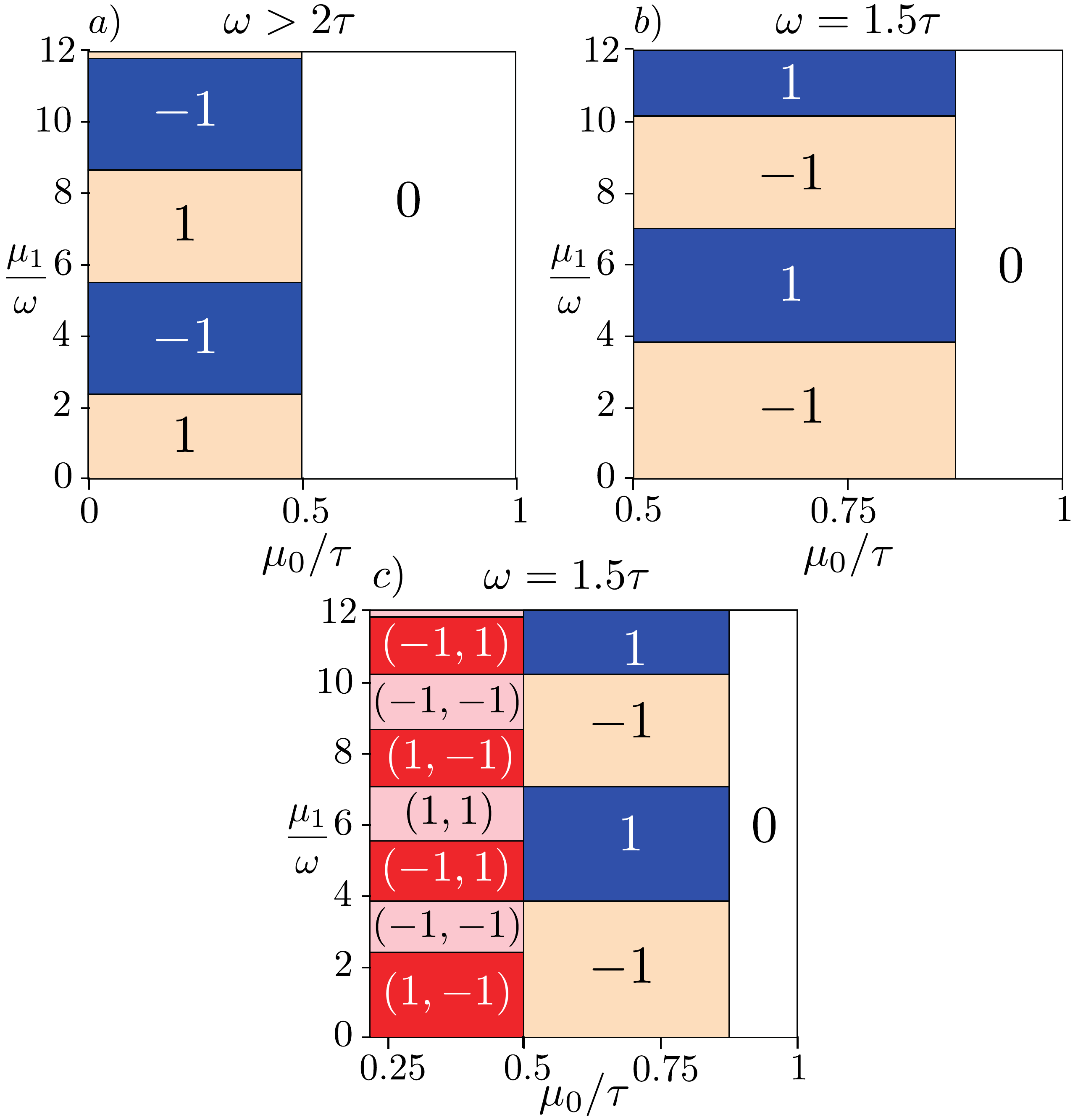}
  \caption{
    (Color online) 
    Phase diagram of the Kitaev Hamiltonian of Eq.~\eqref{eq:kitaev} with time-dependent chemical potential $\mu(t)=\mu_{0}+\frac{\mu_{1}}{2}\cos\omega t$.  The
    white region is topologically trivial ($W=0$), while the other ones are nontrivial ($W\neq0$). (a) In the frequency regime $\omega>2\tau$,  $\tilde{H}_{k}^{\text{eff},0}$
    is used to calculate the bulk invariant $W_0$ for all values of $\mu_{0}$, because
    the series of $\tilde{H}_{k}^{0}(t)$ converges. Transitions between the topological phases $W_{0}=+1$ and $W_{0}=-1$ occur at zeros of ${\cal J}_{0}\left(\frac{\mu_{1}}{\omega}\right)$.
    (b) For $\omega=1.5\tau$, the series of $\tilde{H}_{k}^{1}(t)$ converges
     for $\mu_{0}>0.5\tau$. The trivial phase appears for $\mu_{0}>0.875\tau$ and transitions
     between phases $W_{1}=1$ and $W_{1}=-1$ take place at zeros of ${\cal J}_{1}\left(\frac{\mu_{1}}{\omega}\right)$. (c) Depicts the extension of the phase diagram shown in (b) for $\omega=1.5\tau$ to smaller values of $\mu_{0}$. Besides
     the phases $W_{1}=0,\pm1$, we find new topological phases that are described by two topological
     invariants $(W_{0},W_{1})$, corresponding to the effective Hamiltonians $\tilde{H}_{k}^{\text{eff},0}$ and $\tilde{H}_{k}^{\text{eff},1}$.  We consider a fixed bandwidth $\tau=\mu_{0}+w_{0}$.
    }
    \label{fig:Phase-diagram-high-frequency-lim}
\end{figure}

It is instructive to understand the form of the effective Hamiltonian in Eq.~\eqref{eq:Heffalpha} in terms of Pauli matrices in real space. After a Jordan-Wigner and discrete Fourier transformation we obtain an effective time-independent XY Hamiltonian ~\cite{Lieb1961}
\begin{align}
     \label{eq:effXY}
            \tilde{H}^{\text{eff},\alpha}=&-\frac{\mu_{\text{eff}}}{2}\sum^{N}_{j=1}\sigma^{x}_{j}-\frac{1}{4}\left(w_{0}+\Delta_{\text{eff}}\right)\sum^{N}_{j=1}\sigma^{z}_{j}\sigma^{z}_{j+1}\nonumber \\
            & -\frac{1}{4}\left(w_{0}-\Delta_{\text{eff}}\right)\sum^{N}_{j=1}\sigma^{y}_{j}\sigma^{y}_{j+1} 
      \ .
\end{align}
Apart from the existence of a paramagnetic phase, the effective anisotropies in Eq.~\eqref{eq:effXY} can be tuned to generate a ferromagnetic phase in $z$-direction (FMZ) or $y$-direction (FMY) as it is discussed in Ref.~\cite{Bastidas2012}. In terms of Jordan-Wigner fermions in real space, the effective Hamiltonian Eq.~\eqref{eq:effXY} reads
\begin{align}
\label{eq:kitaeveff} 
 \tilde{H}_{k}^{\text{eff},\alpha}=&\frac{\mu_{\text{eff}}}{2}\sum_{j=1}^{N}\left(2f_{j}^{\dagger}f_{j}-1\right)-\frac{w_0}{2}\sum_{j=1}^{N-1} \left(f_{j}^{\dagger}f_{j+1}+\text{h.c.}\right)
\nonumber\\ &
-\frac{\Delta_{\text{eff}}}{2}\sum_{j=1}^{N-1}\left(f_{j}^{\dagger}f^{\dagger}_{j+1}+\text{h.c.}\right) \ .
\end{align}
The bulk topological invariant is the winding number, which for the effective Hamiltonian 
Eq.~\eqref{eq:Heffalpha} becomes $W_{\alpha}=1/2\pi\int^{2\pi}_{0}d\varphi^{\alpha}_k$, where $\tan\varphi^{\alpha}_k=\Delta_{\text{eff}}\sin k\ (\mu_{\text{eff}}-w_{0}\cos k)^{-1}$. There
is a trivial-nontrivial TQPT at $\mu_{0}-\frac{\alpha\omega}{2}=w_{0}$, where the winding number
changes from $W_{\alpha}=0$ to $W_{\alpha}\neq 0$. In addition, in the nontrivial region, there are
TQPTs between different topological phases at critical lines defined by ${\cal J}_{-\alpha}\left(\frac{\mu_{1}}{\omega}\right)=0$.
The different topological phases are classified by the winding number
$W_{\alpha}=\text{Sign} {\cal J}_{-\alpha}\left(\frac{\mu_{1}}{\omega}\right)$. Fig.~\ref{fig:Phase-diagram-high-frequency-lim}~a) depicts the phase diagram for $\alpha=0$ and Fig.~\ref{fig:Phase-diagram-high-frequency-lim}~b) for $\alpha=1$.

The family of  effective Hamiltonians given by Eq.~\eqref{eq:Heffalpha} can close only at zero quasienergy, while the quasienergies
present also closings at $\epsilon=\frac{\omega}{2}$.  The 
shift of Eq.~\eqref{eq:shift} 
implies that a Hamiltonian
with even $\alpha$ will describe closings of the gap at $\epsilon=0$
and a Hamiltonian with odd $\alpha$ closings of the gap at $\epsilon=\pm\frac{\omega}{2}$.
By now, we have found regions of the parameter space that can be described
with only one effective Hamiltonian $\tilde{H}^{\text{eff},\alpha}_{k}$ for every $k$. Consequently,
in these regions only one of the gaps ($\epsilon=0\ \text{or}\ \frac{\omega}{2}$)
can close. However, it is possible to use two effective Hamiltonians at the same time, e.g. $\tilde{H}^{\text{eff},\alpha}_{k}$ and $\tilde{H}^{\text{eff},\alpha+1}_{k}$, to get a full convergence of the Magnus series in such
a way that one of the Hamiltonians reproduces the $0$-gap and the other
one the $\frac{\omega}{2}$-gap. In the following we probe
that this is possible and show where it can be used and how it works.

\subsection{Combination of frames}
Looking at the convergence
of the Magnus expansion, we realize that even when the series of
$\tilde{H}_{k}^{\alpha}(t)$ does not converge for all $k$-values it
is possible that it converges for some $k$ values while $\tilde{H}_{k}^{\alpha+1}(t)$
converges for the rest. In that case a complete convergence is possible
using both. This concept increases the size of the regions that can
be studied analytically in a good approximation.

This motivates the use of two effective Hamiltonians to classify the topological features of the system. Therefore, we use two invariants
$W_{\alpha}$ and $W_{\alpha+1}$, which give a complete topological
description encoded in the pair $(W_{\alpha},W_{\alpha+1})$. In these cases there are TQPTs at zeros of both ${\cal J}_{\alpha}\left(\frac{\mu_{1}}{\omega}\right)$
and ${\cal J}_{\alpha+1}\left(\frac{\mu_{1}}{\omega}\right)$. 
An example of this situation is shown in Fig.~\ref{fig:Phase-diagram-high-frequency-lim}~c) for $\omega=1.5\tau$,
in which we extend the phase diagram of Fig.~\ref{fig:Phase-diagram-high-frequency-lim}~b) to smaller values of $\mu_{0}$. Fig.~\ref{fig:convergence1} shows that for $\omega>2\tau$ the Magnus series of $\tilde{H}_{k}^{0}(t)$ converges independently of $k$ and $\mu_{0}$. However, for values of $k$ in a neighborhood of $k=0$, the Magnus series converges even for a lower driving frequency $\omega=1.5\tau$ . This allows us to extend the phase diagram of Fig.~\ref{fig:Phase-diagram-high-frequency-lim}~b), because the series of $\tilde{H}_{k}^{1}(t)$ also converges for values in the neighborhood of $k=\pi$ in the region $0.25\tau<\mu_0<0.5\tau$.
In Appendix~\ref{ap-diagram} we discuss the features of the phase diagrams of Fig.~\ref{fig:Phase-diagram-high-frequency-lim} and compare them with the numerical result obtaining a good agreement. Moreover, all the critical lines of the TQPTs are explained in detail from the analytic approach.


\subsection{Majorana end states}

\begin{figure}
\includegraphics[width=1.\columnwidth]{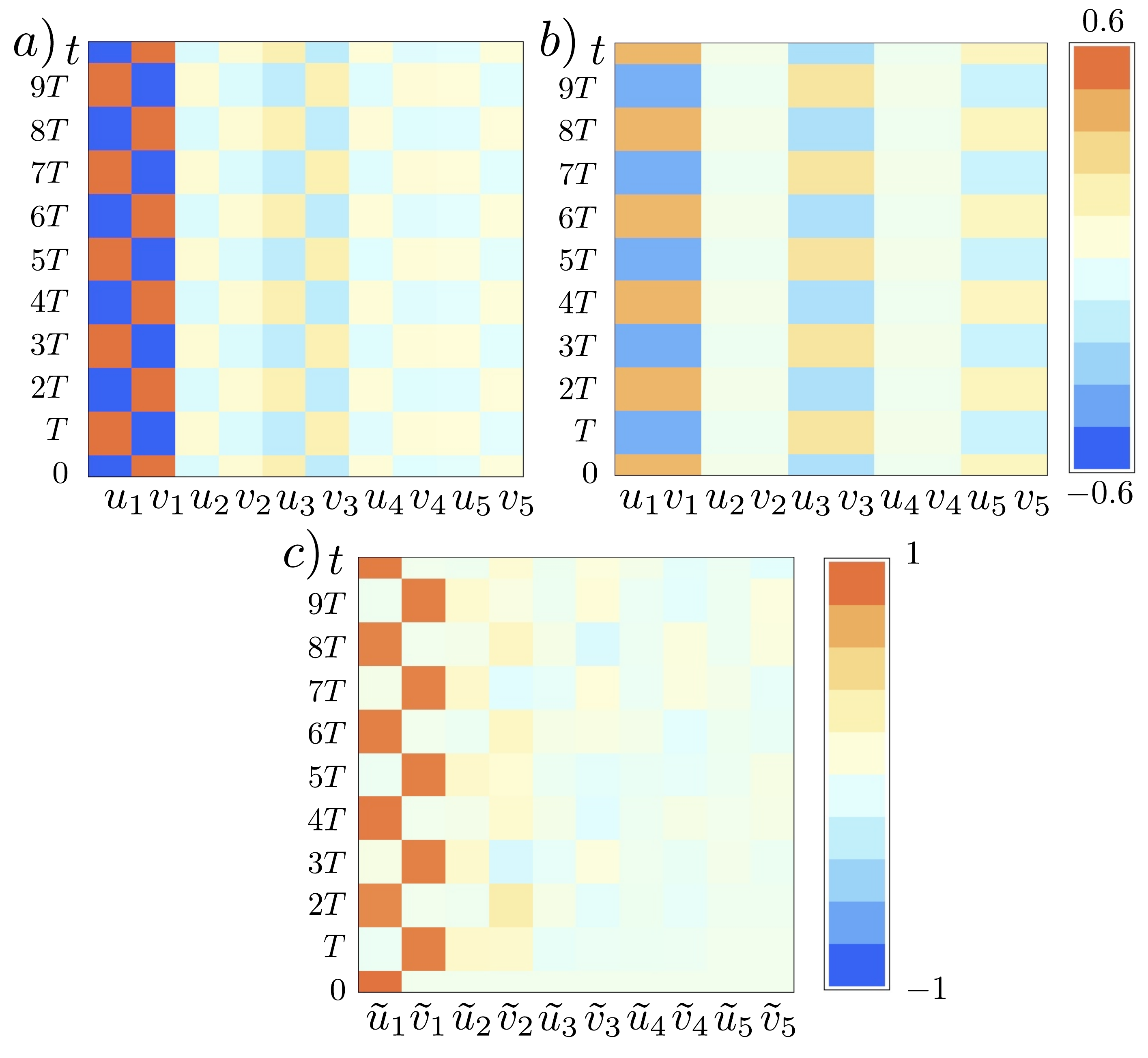}
  \caption{
    (Color online) 
    (a,b) Temporal stroboscopic evolution of the stationary left end state for a finite chain with $N=60$ sites. The color indicates the value of coefficients $u_i$ and $v_i$ in Eq. \eqref{eq:end-states-num} ($l$ correspondent to the left Majorana end mode) for the first 5 sites.  For a frequency $\omega=1.5\tau$, we have performed numerical calculations in the case of a driven chemical potential $\mu(t)=\mu_{0}+\frac{\mu_{1}}{2}\cos\omega t$ with $\mu_0 =0.75\tau$.
    (a) Depicts the evolution for $\frac{\mu_{1}}{\omega}=1$, being ${\cal J}_{-1}\left(\frac{\mu_{1}}{\omega}\right)<0$,
   and (b) for $\frac{\mu_{1}}{\omega}=4.5$, when ${\cal J}_{-1}\left(\frac{\mu_{1}}{\omega}\right)>0$. The stroboscopic dynamic agrees with the predicted  one. (c) Shows the evolution of coefficients $\tilde{u}_i$ and $\tilde{v}_i$  (Eq. \ref{eq:Evol}) in case $\mu_{0}=0.3\tau$ and $\frac{\mu_{1}}{\omega}=1$ imposing
   the initial condition $\Gamma(0)=f_{1}$ at $t=0$. The predicted double period electron-hole oscillations are observed. We consider a fixed bandwidth $\tau=\mu_{0}+w_{0}$.}
    \label{fig:ap1}
\end{figure}

The bulk-boundary correspondence involves the existence of end states localized at the boundary between different bulk topologies. In this section we find the time evolution of the Majorana end state at the boundary between a non-trivial topological phase and the vacuum.

Let us assume that one of the effective Hamiltonians Eq.~\eqref{eq:Heffalpha} 
converges in a particular region of the parameter space, as it is depicted in Fig.~\ref{fig:convergence1}. In the case of open boundary conditions, we can use the Majorana operators \foreignlanguage{american}{$a_{2j-1}=f_{j}+f_{j}^{\dagger}$} and \foreignlanguage{american}{$a_{2j}=-i\left(f_{j}-f_{j}^{\dagger}\right)$} as defined in Ref.~\cite{Kitaev2001} to write the
effective Hamiltonian Eq.~\eqref{eq:kitaeveff} as follows
\begin{align}
\label{eq:kitaev-majoranaop} 
 \tilde{H}_{k}^{\text{eff},\alpha}=&\frac{\mu_{\text{eff}}}{2}\sum_{j=1}^{N}\left(i a_{2j}a_{2j-1}-1\right)
 \nonumber\\ &
 -\frac{i \left(w_0+\Delta_{\text{eff}}\right)}{4}\sum_{j=1}^{N-1} a_{2j}a_{2j+1}
\nonumber\\ &
-\frac{i \left(-w_0+\Delta_{\text{eff}}\right)}{4}\sum_{j=1}^{N-1}a_{2j-1} a_{2j+2} \ .
\end{align}
In the limit $\mu_{\text{eff}}\gg |w_0+\Delta_{\text{eff}}|,|-w_0+\Delta_{\text{eff}}|$  
there are no zero-energy excitations. 
In the case $\mu_{\text{eff}}\ll|w_0-\Delta_{\text{eff}}|$ and $\Delta_{\text{eff}}\simeq -w_0$ the third term dominates and the zero-energy excitations $a_2$ and $a_{2N-1}$ with bulk invariant $W_{\alpha}=-1$ do not appear in the Hamiltonian, but they define the nonlocal fermion $\tilde{f}=\frac{1}{2}(a_2+ia_{2N-1})$, which is topologically protected~\cite{Kitaev2001,Alicea2012}.
In the case of a semi-infinite chain with a large number of sites $N\gg1$, we can obtain time evolution of the left end state in the laboratory frame
\begin{equation}
      \label{eq:end-states-neg}
            \tilde{\gamma}(t)  \approx  -i\left(f_{1}e^{-i\theta_{\alpha}(t)}-f_{1}^{\dagger}e^{i\theta_{\alpha}(t)}\right) 
       \ .
\end{equation}
In the case where $\mu_{\text{eff}}\ll|w_0+\Delta_{\text{eff}}|$ and $\Delta_{\text{eff}}\simeq w_0$, the second term dominates and the Majorana operators $a_1$ and $a_{2N}$ do not appear in the Hamiltonian~\cite{Kitaev2001}.  Similarly to the previous case
they are combined into a nonlocal fermion $f=\frac{1}{2}(a_1+ia_{2N})$. In this regime the system possesses the  bulk invariant $W_{\alpha}=1$ and the time evolution of the mode localized at the first site reads
\begin{equation}
      \label{eq:endstatesJ}
      \gamma(t)  \approx  f_{1}e^{-i\theta_{\alpha}(t)}+f_{1}^{\dagger}e^{i\theta_{\alpha}(t)}
      \ .
\end{equation} 
Interestingly, at discrete times $t=nT$, the edge states of Eqs.~\eqref{eq:end-states-neg} and~\eqref{eq:endstatesJ} are given by $\tilde{\gamma}(nT)\approx-(-1)^{n\alpha}i(f_{1}-f_{1}^{\dagger})$ and $\gamma(nT)\approx(-1)^{n\alpha}(f_{1}+f_{1}^{\dagger})$, respectively.

For a finite chain with $N$ sites,  the numerical calculation explained in Appendix \ref{ap-evolution} allows to obtain the Floquet Majorana modes
\begin{equation}
      \label{eq:end-states-num}
            \boldsymbol{\Psi}_l(t) =  \sum_{i=1}^{N} \left[ u_{li} (t) f_{i}+v_{li} (t) f_{i}^{\dagger}\right]
       \ .
\end{equation}
In order to compare our analytical results with numerical calculations, we consider the Eqs~\eqref{eq:end-states-neg} and~\eqref{eq:endstatesJ} for $\alpha=1$. In this case, the quasienergy gap closes at $\epsilon=\pm\omega/2$. Therefore, to obtain the edge states for a chain with $N=60$ sites, we numerically calculate the
coefficients $u_i(t)$ and $v_i(t)$ for a state with quasienergy $\epsilon=\pm\omega/2$, as we explain in Appendix~\ref{ap-evolution}. By assuming a fixed bandwidth $\tau=\mu_{0}+w_{0}$, we perform the calculation for the parameters $\omega=1.5\tau$ and $\mu_{0}=0.75\tau$.

In order to verify that Eq.~\eqref{eq:end-states-neg} gives us the correct stroboscopic dynamics for $\tilde{\gamma}(nT)$, we plot the imaginary part of the coefficients $u_i(t)$ and $v_i(t)$ in Fig.~\ref{fig:ap1}~a). One can see that at discrete times $t=nT$ they are approximately in agreement with our analytical results. In addition, Fig.~\ref{fig:ap1}~b) depicts the real part of $u_i(t)$ and $v_i(t)$ at discrete times and shows the qualitative agreement with the stroboscopic evolution $\gamma(nT)$ obtained from Eq.~\eqref{eq:endstatesJ}.
The states spread along the vicinity of the end, because the solution of Eqs.~\eqref{eq:end-states-neg}~and~\eqref{eq:endstatesJ}  is only
exact in the limit $w_{0}=\Delta_{0}{\cal J}_{-1}\left(\frac{g_{1}}{\omega}\right)\gg\mu_{0}-\frac{\omega}{2}$, but the weight of the states 
along the chain decreases exponentially.

In a case with end states in the gaps $\epsilon=0$ and $\epsilon=\omega/2$ 
we require the use of the effective Hamiltonians $\tilde{H}_{k}^{\text{eff},\alpha}$ and $\tilde{H}_{k}^{\text{eff},\alpha+1}$. For instance, in the phase with the invariant $\left(W_{0},W_{1}\right)=\left(1,-1\right)$,
there are two Majorana end states, one in each gap
\begin{equation}
\label{eq:dos-estados}
\begin{split}
\gamma_{\epsilon=0}(t) & \approx f_{1}e^{-i\theta_{0}(t)}+f_{1}^{\dagger}e^{i\theta_{0}(t)}\ ,\\
\gamma_{\epsilon=\frac{\omega}{2}}(t)  & \approx  -i\left(f_{1}e^{-i\theta_{1}(t)}-f_{1}^{\dagger}e^{i\theta_{1}(t)}\right) \ , 
\end{split}
\end{equation}
for parameters $\omega=1.5\tau$, $\mu_{0}=0.3\tau$ and $\mu_{1}/\omega=1$.  
The fact that two non-degenerate (in quasienergy) end states are present in the system generates interferences characteristic of ac-driven topological systems~\cite{Jiang2011,Kitagawa2012}.
In order to see the interference of states in both gaps we are interested in the study of the time evolution of the system for a given initial condition. 
According to the approximated Majorana modes, if the initial excitation is $\Gamma(0)=f_{1}$ it can be written as $\Gamma(0)\approx\left[\gamma_{\epsilon=0}(0)+i\gamma_{\epsilon=\frac{\omega}{2}}(0)\right]/2$.
Therefore, the evolved excitation
is known to be $\Gamma(t)\approx\left[\gamma_{\epsilon=0}(t)+i\gamma_{\epsilon=\frac{\omega}{2}}(t)\right]/2$ at all times. At discrete times $t=nT$ the system exhibits a doubly-periodic stroboscopic dynamics $\Gamma(nT)\approx\frac{f_{1}}{2}\left[1+(-1)^n\right]+\frac{f_{1}^{\dagger}}{2}\left[1-(-1)^n\right]$.
In Appendix~\ref{ap-evolution} we explained how to obtain the evolved excitation after a imposed initial condition, written as 
\begin{equation}
  \label{eq:Evol}
            \boldsymbol{\Gamma}(t)=\sum^{N}_{i=1}[\tilde{u}_{i}(t)f_{i}+\tilde{v}_{i}(t)f^{\dagger}_{i}]
      \ .
\end{equation}
We show the predicted doubly periodic oscillations in  Fig.~\ref{fig:ap1}~c),  where we plot the real part of the numerically obtained coefficients $\tilde{u}_i(t)$ and $\tilde{v}_i(t)$ of $\Gamma(t)$ in Eq. \ref{eq:Evol} at discrete times, for the initial condition $\Gamma(0)=f_{1}$. These oscillations are due to the interference of states in both gaps.

\section{Driving the Tunneling and BCS Pairing\label{SectionIV}}
In this section we apply the findings of last section to study
the Kitaev model considering a different driving protocol, which introduces
different phases with more end states. We consider the Kitaev chain
of spinless fermions $f_{j}$  given in Eq.~\eqref{eq:kitaev}
with a constant chemical potential $\mu_{0}$ and time periodic tunneling and BCS
pairing such that $w(t)=\Delta(t)=J(t)=J_{0}+\frac{J_{1}}{2}\cos\omega t$. Using the equivalence
of Kitaev and Ising model and a duality transformation, the resolution
of the problem is straightforward.
\begin{figure}
\includegraphics[width=1.\columnwidth]{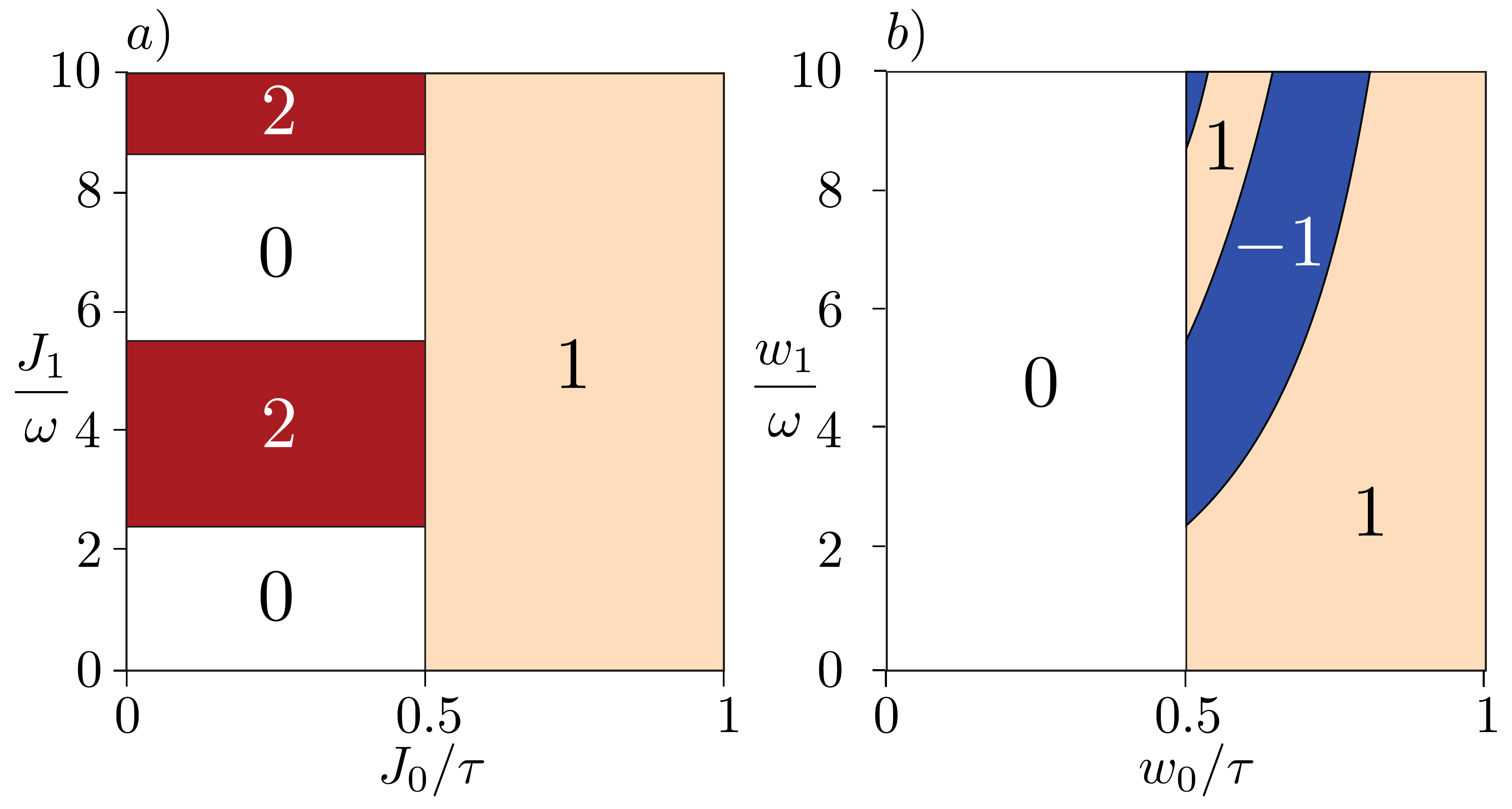}

\caption{Phase diagram for $\omega>2\tau$ in the case of $\mu(t)=\mu_0$, a) $\Delta(t)=w(t)=J(t)=J_{0}+\frac{J_{1}}{2}\cos\omega t$, b) $w(t)=w_{0}+\frac{w_{1}}{2}\cos\omega t$ and $\Delta(t)=\Delta_0$. White region
is topologically trivial ($W=0$), light-orange ($W=1$) and blue ($W=-1$) are nontrivial phases with
one end-state and brown region ($W=2$) is a nontrivial  phase with two end states.  \label{fig:Phase-diagram-high-frequency-lim-driving2}}
\end{figure}

After a Jordan-Wigner transformation, this model corresponds
to the one-dimensional Ising model in an external magnetic field
\begin{equation}
H(t)=-\frac{\mu_0}{2}\sum_{j=1}^{N}\sigma_{j}^{x}-\frac{J(t)}{2}\sum_{j=1}^{N}\sigma_{j}^{z}\sigma_{j+1}^{z}\ ,\label{eq:ising2}
\end{equation}
which follows directly from Hamiltonian Eq.~\eqref{eq:ising-1} with $\mu(t)=\mu_{0}$, $J_{z}(t)=J(t)$ and $J_{y}(t)=0$. Under the duality transformation $   \sigma_{i}^{x} =\mu_{i}^{z}\mu_{i+1}^{z}$, $\sigma_{i}^{z}=\Pi_{k\leq i}\mu_{k}^{x}$~\cite{PhysRevB.88.165111}
we get what is called the dual Hamiltonian of Eq.~\eqref{eq:ising2}
\begin{equation}
      \label{eq:DualIsing}
     H^{(\text{D})}(t)=-\frac{J(t)}{2}\sum_{j=1}^{N}\mu_{j}^{x}-\frac{\mu_{0}}{2}\sum_{j=1}^{N}\mu_{j}^{z}\mu_{j+1}^{z}\ \ ,
\end{equation}
which is exactly the corresponding Ising model to the system studied
in the last section with $\mu(t)\rightarrow J(t)$
and $\Delta(t)=w(t)\rightarrow \mu_{0}$ in Eq.~\eqref{eq:ising-1}---in this case, however, written in terms
of the Pauli matrices $\mu^{\lambda}_i$.
It means that the quasienergy spectrum
is the same, and by performing the inverse duality transformation to the
effective Hamiltonians in Eq.~\eqref{eq:effXY}, we obtain the
effective Hamiltonians for the new driven system Eq.~\eqref{eq:ising2}. 
The effective Hamiltonian in spin basis reads 
\begin{eqnarray}
\tilde{H}^{\text{eff}}_{\alpha} & = & -\frac{J_{\text{eff}}}{2}\sum_{j=1}^{N}\sigma_{j}^{z}\sigma_{j+1}^{z}-\frac{\mu_{0}}{4}\left[1+{\cal J}_{-\alpha}\left(\frac{J_{1}}{\omega}\right)\right]\sum_{j=1}^{N}\sigma_{j}^{x}\nonumber \\
 &  & +\frac{\mu_{0}}{4}\left[1-{\cal J}_{-\alpha}\left(\frac{J_{1}}{\omega}\right)\right]\sum_{j=1}^{N}\sigma_{j-1}^{z}\sigma_{j}^{x}\sigma_{j+1}^{z}\ ,\label{eq:cluste}
\end{eqnarray}
where $J_{\text{eff}}=J_{0}-\frac{\alpha\omega}{2}$. The last term is a three-spins interaction, which in the spinless fermions basis is a second-neighbors
interaction term that will give rise to a new topological phase with winding number $W_{\alpha}=2$.

The phase diagrams for the present configuration are like the ones found
in the previous section (Fig.~\ref{fig:Phase-diagram-high-frequency-lim}), but with a different value of the topological invariants. The phase diagram in the high-frequency regime is shown in Fig.~\ref{fig:Phase-diagram-high-frequency-lim-driving2} a).
It is important to notice that the trivial region in the static case
($J_{1}=0$), becomes nontrivial 
in the high-frequency regime. In Fig.~\ref{fig:Phase-diagram-high-frequency-lim-driving2} a) the bulk invariant can be tunned from trivial ($W_{\alpha}=0$) to nontrivial ($W_{\alpha}=2$), depending on strength of the driving. In contrast to this, in the case of a driven chemical potential, the trivial region remains trivial at high frequency independently of the strength of the driving.

In the case $J_{\text{eff}}>\mu_0$, the chain supports end states given in Eq.~\eqref{eq:endstatesJ}, while no end states are present if $J_{\text{eff}}<\mu_0$ and ${\cal J}_{-\alpha}\left(\frac{J_{1}}{\omega}\right)>0$. To study the existence of end states in the new phase ($J_{\text{eff}}<\mu_0$ and ${\cal J}_{-\alpha}\left(\frac{J_{1}}{\omega}\right)<0$), we write the second-neighbors interaction term in terms of Majorana operators~\cite{Kitaev2001}
\foreignlanguage{american}{$a_{2j-1}=f_{j}+f_{j}^{\dagger}$}
and \foreignlanguage{american}{$a_{2j}=-i\left(f_{j}-f_{j}^{\dagger}\right)$} as \foreignlanguage{american}{$H\propto i\sum_{j}a_{2j-2}a_{2j+1}$}.
Therefore, if the chain is semi-infinite the Majorana operators $a_{1}$ and $a_{3}$ will not appear in the Hamiltonian, being therefore two Majorana end states:
\foreignlanguage{american}{
\begin{equation}
\label{eq:endstatesJ2}
\begin{split}
\gamma_a(t) & \simeq f_{1}e^{-i\theta_{\alpha}(t)}+f_{1}^{\dagger}e^{i\theta_{\alpha}(t)} \ , \\
\gamma_b(t) &\simeq f_{3}e^{-i\theta_{\alpha}(t)}+f_{3}^{\dagger}e^{i\theta_{\alpha}(t)} \ .
\end{split}
\end{equation}
}
To sum up, we have found an effective Kitaev model with second neighbors tunneling and BCS pairing and, 
consequently, a topological phase hosting two Majorana end-states.

\section{Driving the Tunneling\label{SectionV}}

Finally, and for completeness, we are interested in the consequences
of a driving just of the tunneling term of the Kitaev model Eq.~\eqref{eq:kitaev}. We consider a constant chemical potential $\mu_{0}$, 
BCS pairing $\Delta_{0}$, and a monochromatic driving of the tunneling strength $w(t)=w_{0}+\frac{w_{1}}{2}\cos\omega t$.
\begin{figure}
\includegraphics[width=0.85\columnwidth]{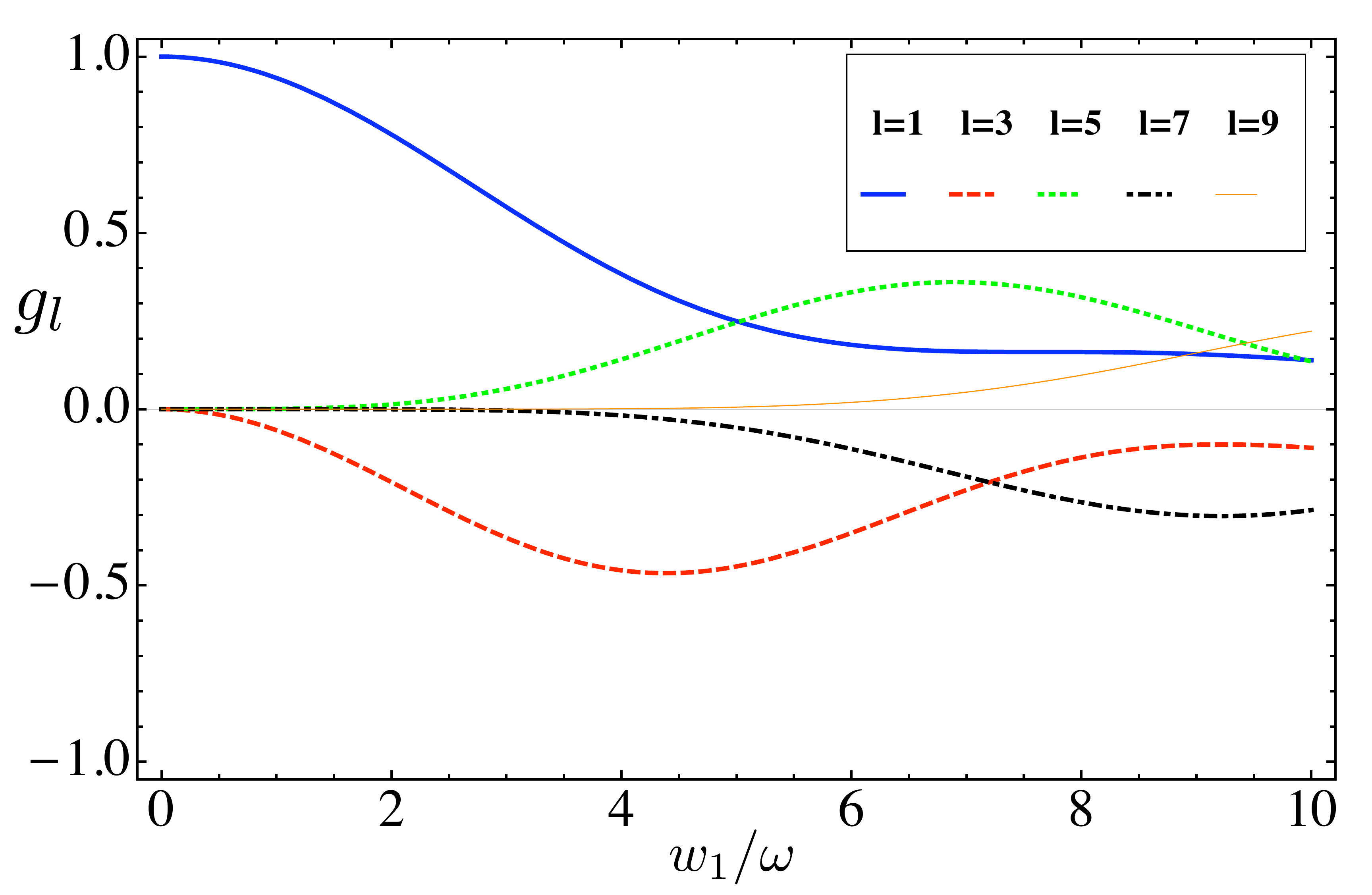}
\caption{BCS interaction strength $g_l$ as a function of $w_1/\omega$ for neighbors $l=1,3,5,7,9$. Notice that the sign of the interaction is $(-1)^{\frac{l-1}{2}}$.  \label{fig:funcionesgl}}
\end{figure}
In this section we obtain a solution in the high-frequency limit by means of a transformation into the interaction picture $\mathcal{S}_{k}=e^{i\theta(t)\cos k\sigma^{z}_{k}}$, where $\theta(t)=\frac{w_{1}}{2\omega}\sin\omega t$. The high-frequency effective Hamiltonian is
\begin{equation}
      \label{eq:5}
            \tilde{H}_{k}^{\text{eff}}=(\mu_{0}-w_{0}\cos k)\sigma^{z}_{k}+\Delta_{0}\sin k\ {\cal J}_{0}\left(\frac{w_{1}}{\omega}\cos k\right)\sigma^{y}_{k}
      \ .
\end{equation}
The transition between trivial and nontrivial phases takes
place at $\mu_{0}=w_{0}$. More gap closings are found when ${\cal J}_{0}\left(\frac{w_{1}}{\omega}\cos k\right)=0$
and $\mu_{0}-w_{0}\cos k=0$. This implies ${\cal J}_{0}\left(\frac{w_{1}}{\omega}\frac{\mu}{w_{0}}\right)=0$,
as long as $\mu_{0}<w_{0}$. 
The high-frequency phase diagram is shown in Fig.~\ref{fig:Phase-diagram-high-frequency-lim-driving2}~b).

Despite its apparent simplicity, the Bogoliubov Hamiltonian Eq.~\eqref{eq:5} encloses a rich physical meaning. The first term of Eq.~\eqref{eq:5} is trivial in the sense that it describes the local term proportional to the chemical potential $\mu_{0}$, and the tunneling between nearest neighbors with amplitude $w_{0}$. The second term, however generates new features of the BCS pairing, which arise from effective long-range interactions in real space.
To understand the nature of these interactions arising in Eq.~\eqref{eq:5}, let us consider the BCS term
\begin{align}
      \label{eq:besselHam}
            V&=
            \frac{\Delta_0}{2i}\sum_{k}\sin k\ {\cal J}_{0}\left(\frac{w_{1}}{\omega}\cos k\right)
            \left(f_{-k}f_{k}-f^{\dagger}_{k}f^{\dagger}_{-k}\right)
      .
\end{align}
Performing the inverse Fourier transformation in order to obtain the real space representation is not straightforward due to the $k$-dependence in the argument of the Bessel function. However, we can use the expansion of the Bessel function in power series of its argument
\begin{align}
            {\cal J}_{0}\left(z_{k}\right)&=\sum_{m=0}^{\infty}\frac{(-1)^{m}}{m!^2} \left(\frac{w_{1}}{2\omega}\cos k\right)^{2m} \label{eq:BesselSeries} \\
            &=\sum_{m=0}^{\infty}\sum^{2m}_{r=0}\frac{(-1)^{m}}{m!^2}\left(\frac{w_{1}}{4\omega}\right)^{2m}\begin{pmatrix}2m \\ r\end{pmatrix}e^{2i(m-r)k} \nonumber
       \ ,
\end{align}
where the binomial theorem was used. After some manipulations we get
\begin{equation}
\begin{split}
V=-\frac{\Delta_{0}}{2}\sum_{j=1}^{N}\sum^{\infty}_{m=0}\sum^{2m}_{r=0}C_{m,r} & \left (  f_{j}^{\dagger}f_{j+a_{m,r}}^{\dagger}\right. \label{eq:EffBCS} \\ & \left. - f_{j}^{\dagger}f_{j+b_{m,r}}^{\dagger}+h.c.\right)  ,
\end{split}
\end{equation}
with coefficients 
\begin{equation}
      \label{eq:coeffExp}
            C_{m,r}=\frac{(-1)^{m}}{2(m!)^2}\left(\frac{w_{1}}{4\omega}\right)^{2m}
            \begin{pmatrix}2m \\ r\end{pmatrix} \ .
\end{equation}
This means that $\text{V}$ is an effective BCS interaction term with neighbors range given by
\begin{equation}
 \label{eq:longRange}
\begin{split}
            a_{m,r}&=2(m-r)+1 \ , \\
            b_{m,r}&=2(m-r)-1 \ .
\end{split}
\end{equation}
Finally, the Hamiltonian in real space can be simplified to 
\begin{align}
\label{eq:EffBCSsimp}
\tilde{H}^{\text{eff}}  = &  \frac{\mu_{0}}{2}\sum_{j=1}^{N}\left(2f_{j}^{\dagger}f_{j}-1\right)-\frac{w_{0}}{2}\sum_{j=1}^{N}\left( f_{j}^{\dagger}f_{j+1}+h.c.\right) \nonumber \\
 &   -\frac{\Delta_{0}}{2}\sum_{j=1}^{N}\sum^{N}_{l=1,3...}g_l(w_1)\left(f_{j}^{\dagger}f_{j+l}^{\dagger}+h.c.\right) \  ,
\end{align}
where the strength of the BCS interaction between l-th neighbors is given by the function $g_l(w_1)$:
\begin{align}
\label{eq:functiong}
g_l(w_1) & =  2\left[D\left(\frac{l-1}{2}\right)-D\left(\frac{l+1}{2}\right)\right]  \nonumber \ , \\
D(d) & = \sum_{m=0}^{\infty}C_{m,m-d} \ .
\end{align}
We show the function $g_l(w_1)$ in Fig.~\ref{fig:funcionesgl}. As expected, for small $w_1$ the first-neighbors interaction is larger. However, as the amplitude of the driving increases the next-neighbors interactions become important. Since the sign between the first-neighbors and third-neighbors interaction is opposite  (Fig.~\ref{fig:funcionesgl}), the winding number changes sign as it was shown in Fig.~\ref{fig:Phase-diagram-high-frequency-lim-driving2}~b).

In the reciprocal space, the Hamiltonian Eq.~\eqref{eq:EffBCSsimp} reads
\begin{equation}
\label{eq:sinksin3k}
            \tilde{H}_{k}^{\text{eff}}=(\mu_{0}-w_{0}\cos k)\sigma^{z}_{k}+\Delta_{0}\sum^{N}_{l=1,3...}g_l(w_1)\sin{kl} \ \sigma^{y}_{k}
      \  ,
\end{equation}
which shows how the driving allows to engineer exotic phases of matter. 
Finally, we derive the expression of the long-range BCS interaction arising in Eq.~\eqref{eq:EffBCSsimp} in terms of Pauli matrices in real space. This allows us to obtain the effective spin model 
\begin{align}
\tilde{H}^{\text{eff}} & =  -\frac{\mu_0}{2}\sum_{j=1}^{N}\sigma_{j}^{x}-\frac{w_{0}}{4}\sum_{j=1}^{N}\left(\sigma_{j}^{z}\sigma_{j+1}^{z}+\sigma_{j}^{y}\sigma_{j+1}^{y}\right)  \\
 &   -\frac{\Delta_{0}}{4}\sum_{j=1}^{N}\sum_{l=1,3...}^{N} g_l(w_1) \left(\sigma_{j}^{z} M^{x}_{j,l}\sigma_{j+l}^{z}-\sigma_{j}^{y} M^{x}_{j,l} \sigma_{j+l}^{y}\right) \nonumber \ ,\label{eq:prodpauli}
\end{align}
where $M^{x}_{i,l}=\sigma_{i+1}^{x}...\sigma_{i+l-1}^{x}$. Long-range  spin interactions are generated by means of the ac driving of the tunneling, which in the spins basis corresponds to a time-periodic anisotropy between Z and Y directions (see Eq~\ref{eq:ising-1}).

\

The Kitaev chain is a simple model that considers spinless fermions.
A physical realization of this model is a one-dimensional wire with
Rashba spin-orbit interaction, Zeeman splitting and proximity induced
s-wave superconductivity~\cite{Lutchyn2010}. 
In this realization, the periodic variation of the chemical potential in the wire is possible by means of an alternating gate voltage applied to the substrate, as suggested in ~\cite{Wu2013}. 
Another proposed realization
of the Kitaev chain consists in using semiconductor quantum dots coupled
to superconducting grains~\cite{2012NatCo...3E.964S}. In this setup the access to the other parameters is more suitable because the relations between the experimental and effective parameters ir more simple~\cite{Fulga2013}. The advantage
of our analytical approach is that it allows to easily predict the
TQPTs at any frequency regime i.e., not only in the high frequency regime but also at intermediate and low frequencies. 
Our approach allows the comparison with future experiments performed into a full range of frequency regimes of the external driving
as far as we restrict ourselves to the studied convergence regions.

\section{Conclusions\label{SectionVI}}

We have discussed the nonequilibrium TQPTs in the Kitaev model with
three different driving protocols. In all the cases, we focus on the effect of monochromatic
control of the parameters, which is a realistic driving. 
By means of rotations of
frame, we get a completely analytical description of the topological phase diagram in a wide range of frequencies for some values of the parameters.
Moreover, we are able to provide an approximated wave function of the Majorana end states.

The equivalence between Kitaev model and Ising model allows to use
a simple duality transformation to relate the previous results with
the resolution of a Kitaev chain whose tunneling and BCS parameters
are varied in time harmonically. In this case, new features are found, like the appearance of two Majorana end states at high frequency.

Finally,
by only driving the tunneling
, very interesting effective models with  long-range superconductivity arise.
Our analysis addressing
three different ways of driving with harmonic time dependent potentials
gives a full picture of the consequences of the topological phases at arbitrary
frequencies. It allows to design the most efficient way to search
signatures of Floquet Majorana fermions by appropriated drive of the system. Moreover, we briefly explain the equivalence of these properties in the spin chain basis, emphasizing the novelties detected.

\begin{acknowledgments}
M.B., A.G-L and G.P acknowledge the Spanish Ministry
of Economy and Competitiveness through grant no.
MAT2011-24331 and the associated FPI scholarship (M.B.).
V. M. B. and T. B. wish to acknowledge financial
support by the DFG via GRK 1558, grants BRA 1528/7, BRA 
1528/8, SFB 910.
\end{acknowledgments}

\appendix

\section{Explicit description of the combination of frames\label{ap-diagram}}
\begin{figure}
\includegraphics[width=1\columnwidth]{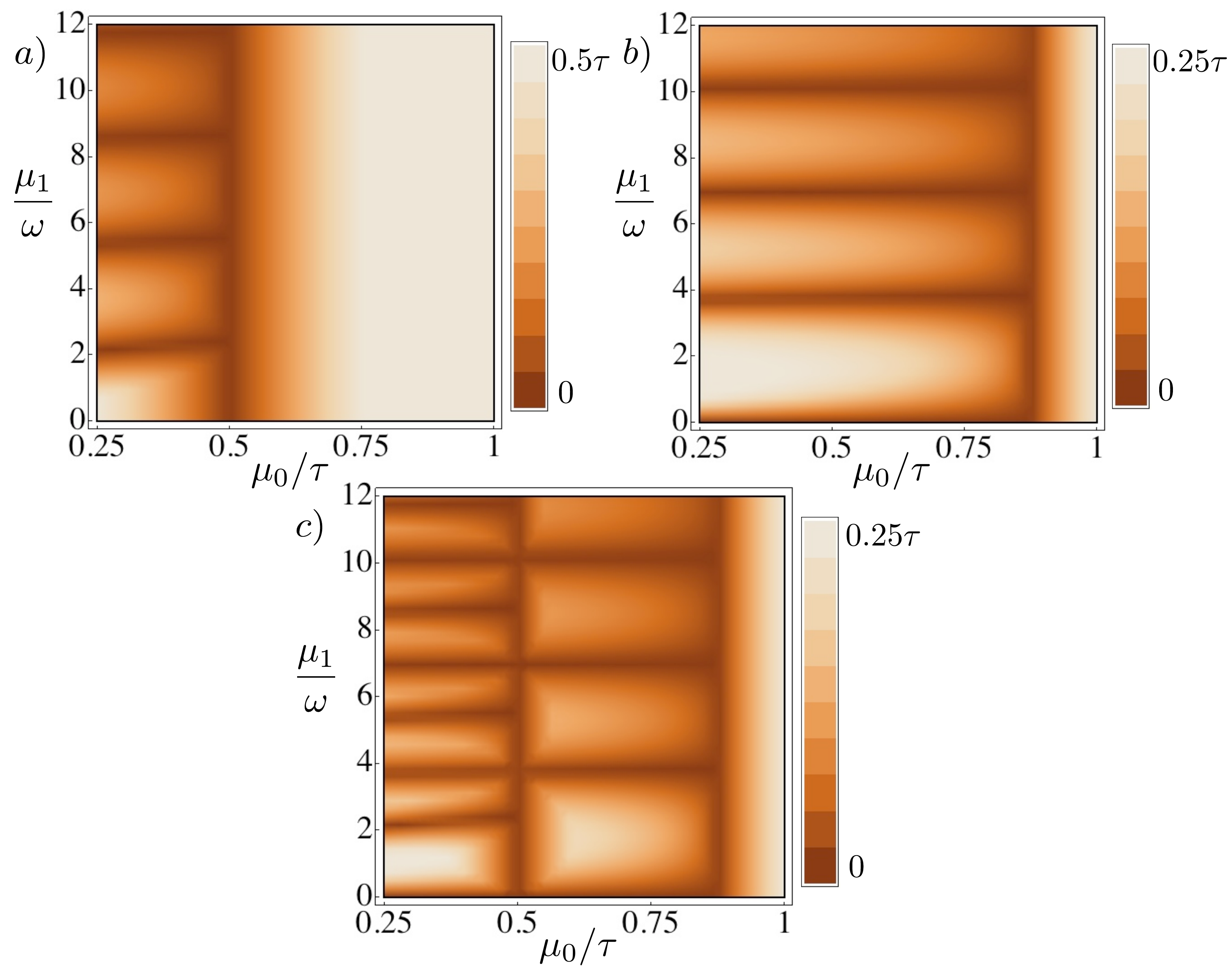}

\caption{Phase diagram for $\omega=1.5\tau$ in the case of a driven
chemical potential. The dark lines show the closings of the:
a) $0-$gap, b) $\frac{\omega}{2}$-gap, and c) both gaps. The whole topological phase diagram is given by c), which agrees
perfectly with the analytical result in Fig.~ \ref{fig:Phase-diagram-high-frequency-lim}. \label{fig:Phase-diagram-1p5}}
\end{figure}
In this appendix we discuss the phase diagram of  Fig.~\ref{fig:Phase-diagram-high-frequency-lim}  for $\omega=1.5\tau$. We analyze in  more detail the origin of each TQPT and calculate the critical lines numerically in order to check the validity of our approximated method used in section \ref{SectionIII}. In Fig~\ref{fig:Phase-diagram-1p5}~a) the critical lines corresponding to the $0$-gap of quasienergies are shown, while the critical lines of the $\frac{\omega}{2}$-gap are represented in Fig.~\ref{fig:Phase-diagram-1p5}~b). The dark lines indicate the occurrence of TQPTs. The combination of both gaps provides the whole topological phase diagram depicted in 
Fig.~\ref{fig:Phase-diagram-1p5}~c). 
\begin{figure}[h]
\includegraphics[width=1\columnwidth]{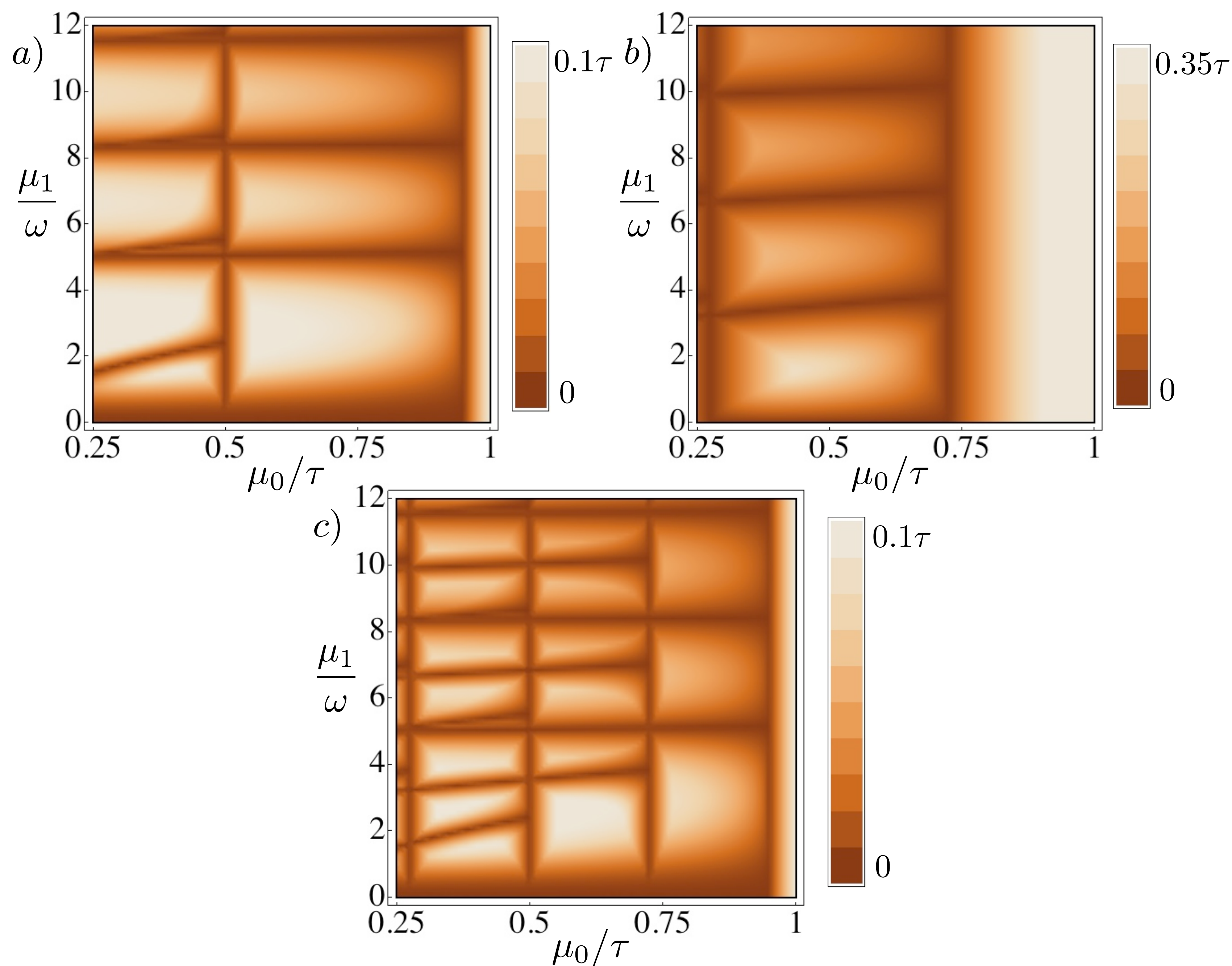}

\caption{Phase diagram for $\omega=0.9\tau$ in the case of a driven
chemical potential. The dark lines show the closings of the:
a) $0-$gap, b) $\frac{\omega}{2}$-gap and c) both gaps. The diagram
c) in the right part is the same as diagram for $\omega=1.5\tau$
but changing the zeros of ${\cal J}_{0}$ by zeros of ${\cal J}_{1}$
and the zeros of ${\cal J}_{1}$ by zeros of ${\cal J}_{2}$.  ($\tau=1$). \label{fig:Phase-diagram-0p9}}
\end{figure}

By means of the method developed
in the main text, we are able to explain the different phases present
in this phase diagram. In the regime $\tau<\omega<2\tau$ and for any value of $k$, the convergence condition
of Eq.~\eqref{eq:convNambu} for $\tilde{H}_{k}^{\alpha=1}(t)$
can be reduced to $\mu_{0}>0.5\tau$. On the other hand, the 
trivial-nontrivial transition occurs at $\mu_{0}-\frac{\omega}{2}=w_0$, which implies
$\mu_{0}=\frac{\tau}{2}+\frac{\omega}{4}$. By fixing the frequency, $\omega=1.5\tau$, we predict TQPTs at zeros of the Bessel function ${\cal J}_{1}\left(\frac{\mu_{1}}{\omega}\right)$, which occur at values $\frac{\mu_{1}}{\omega}\in\{3.8, 7.0, 10.2 ...\}$, in the regime $0.5\tau<\mu_{0}<0.875\tau$ shown in Fig. \ref{fig:Phase-diagram-1p5}~b). For smaller values
of $\mu_{0}$, we need to use $\tilde{H}_{k}^{\alpha=0}(t)$ and $\tilde{H}_{k}^{\alpha=1}(t)$, as we explained in section~ \ref{SectionII},
and consequently we predict TQPTs at zeros of ${\cal J}_{0}\left(\frac{\mu_{1}}{\omega}\right)$
and ${\cal J}_{1}\left(\frac{\mu_{1}}{\omega}\right)$ as in 
Fig.~\ref{fig:Phase-diagram-1p5}~c).

We show also in Fig.~\ref{fig:Phase-diagram-0p9} the same calculation
for $\omega=0.9\tau$, in order to see that for large values
of $\mu_{0}$ the analytical approach is useful. For $\frac{2}{3}\tau<\omega<\tau$,
the Magnus series of $\tilde{H}_{k}^{\alpha=2}(t)$ converges if $\mu_{0}>\frac{\omega}{4}+\frac{\tau}{2}$.
For $\omega=0.9\tau$, this value is $\mu_{0}>0.725\tau$. On the other hand,
the trivial-nontrivial transition occurs at $\mu_{0}-\omega=w_0$.
This means $\mu_{0}=\frac{\tau}{2}+\frac{\omega}{2}=0.95\tau$. Then, for values
$0.725\tau<\mu_{0}<0.95\tau$, the phase diagram shows TQPTs at
zeros of ${\cal J}_{2}\left(\frac{\mu_{1}}{\omega}\right)$, 
which appear at values $\frac{\mu_{1}}{\omega}\in\{5.1,
8.4, 11.6, ...\}$, as it is shown in Fig.~\ref{fig:Phase-diagram-0p9}~c).
For smaller values of $\mu_{0}$, we predict TQPTs at zeros of ${\cal J}_{2}\left(\frac{\mu_{1}}{\omega}\right)$
and ${\cal J}_{1}\left(\frac{\mu_{1}}{\omega}\right)$. However, for even smaller values of $\mu_{0}$, our analytical
approach is not valid anymore and the phase diagram is more complex.  

\section{Numerical calculation of temporal evolution\label{ap-evolution}}
In this appendix we want to give more details about the numerical calculation of the temporal evolution shown in Fig.~\ref{fig:ap1}. Following the seminal paper of Kitaev~\cite{Kitaev2001}, one can write the Hamiltonian Eq.~\eqref{eq:kitaev} as a quadratic form in terms of the Majorana operators $a_{2j}$ and $a_{2j-1}$
\begin{equation}
      \label{eq:KitaevQuadMajorana}
            H(t)=\frac{i}{4}\sum^{2N}_{l,r}a_{l}M_{lr}(t)a_{r}
      \ ,
\end{equation}
where $M(t)$ is a time-periodic real antisymmetric matrix. Motivated by a previous work~\cite{Thakurathi2013}, we calculate the Heisenberg equations of motion for the Majorana operators 
\begin{equation}
      \label{eq:HeisenbergMajorana}
            \frac{d a_{l}(t)}{d t}=i[H(t),a_{l}(t)]=\sum^{2N}_{r=1}M_{lr}(t)a_{r}(t)
      \ .
\end{equation}
Following Refs.~\cite{Kitaev2001,Thakurathi2013} we construct the column vector $\boldsymbol{A}=(a_1,a_2,\ldots,a_{2N})^{\text{T}}$, which allows us to write Eq.~\eqref{eq:HeisenbergMajorana} as follows
\begin{equation}
      \label{eq:VectorHeisenbergMajorana}
            \frac{d \boldsymbol{A}(t)}{d t}=M(t)\boldsymbol{A}(t)
      \ .
\end{equation}
Now we can build the evolution operator such that $\boldsymbol{\mathcal{U}}(t,0)\boldsymbol{A}(0)=\boldsymbol{A}(t)$. 
Invoking Floquet theory~\cite{Shirley1965} we look for
Floquet states of the form $\boldsymbol{\Psi}_{l}(t)=e^{-i\epsilon_{l}t}\boldsymbol{\Upsilon}_{l}(t)$ with Floquet modes
$\boldsymbol{\Upsilon}_{l}(t)=\boldsymbol{\Upsilon}_{l}(t+T)$ satisfying
the eigenvalue equation $\boldsymbol{\mathcal{U}}(T,0)\boldsymbol{\Upsilon}_{l}(0)=e^{-i\epsilon_{l}T}\boldsymbol{\Upsilon}_{l}(0)$.
Our aim is to obtain the Majorana field operators corresponding to the end state with quasienergies $\epsilon_{l}=0,\pm\frac{\omega}{2}$. In so doing we just need to find the corresponding $\boldsymbol{\Upsilon}_{\tilde{l}}(t)$ such that $\boldsymbol{\mathcal{U}}(T,0)\boldsymbol{\Upsilon}_{\tilde{l}}(0)=\boldsymbol{\Upsilon}_{\tilde{l}}(0)$ for the $\epsilon_{\tilde{l}}=0$ gap, and $\boldsymbol{\mathcal{U}}(T,0)\boldsymbol{\Upsilon}_{\tilde{l}}(0)=e^{\mp i \pi}\boldsymbol{\Upsilon}_{\tilde{l}}(0)$ for the $\epsilon_{\tilde{l}}=\pm\frac{\omega}{2}$ gap. These Majorana field operators can be written 
in terms of complex fermions
\begin{equation}
      \label{eq:MajoranaFieldZero}
             \boldsymbol{\Psi}_{\tilde{l}}(t)= \sum_{r=1}^{N} \left[ u_{\tilde{l}r}(t) f_{r}+v_{\tilde{l}r} (t) f_{r}^{\dagger}\right]
      .
\end{equation}
In the main text, for simplicity, we have dropped the index $l$ in the definition of the Majorana mode of Eq.~\eqref{eq:MajoranaFieldZero}.
The coefficients $u_i(t)$ and $v_i(t)$ are shown in Figs.~\ref{fig:ap1} a) and b) for different cases and they are in agreement with the predicted ones. 

Given a initial condition $\boldsymbol{\Gamma}(0)=f_1=(a_1(0)+i a_2(0))$ the evolution is known to be $\boldsymbol{\Gamma}(t)=(a_1(t)+i a_2(t))$, written in a general form as:
\begin{equation}
  \label{eq:FermionEvol}
            \boldsymbol{\Gamma}(t)=\sum^{N}_{r=1}[\tilde{u}_{r}(t)f_{r}+\tilde{v}_{r}(t)f^{\dagger}_{r}]
      \ .
\end{equation}

\bibliographystyle{phaip}
\bibliography{MyCollection}

\end{document}